\documentclass[aps,prl,twocolumn,superscriptaddress,nofootinbib,amsmath,amssymb,preprintnumbers]{revtex4-2}
\usepackage{bm}
\usepackage{graphicx}
\usepackage{booktabs}
\usepackage{xcolor}
\usepackage{enumitem}
\usepackage{hyperref}
\hypersetup{colorlinks=true,linkcolor=blue!60!black,citecolor=blue!60!black,
            urlcolor=blue!60!black}
\newcounter{suppsec}
\renewcommand{\thesuppsec}{S\arabic{suppsec}}
\newcommand{\suppsection}[2]{\par\refstepcounter{suppsec}\label{#1}\section*{\thesuppsec. #2}}

\newcommand{\Kc}{\mathcal K}
\newcommand{\nin}{\hat n_{\rm in}}
\newcommand{\nout}{\hat n_{\rm out}}
\newcommand{\vp}{\varpi}

\begin{document}
\preprint{LAPTH-050/26} 
\title{Horizon exchange and the universal all-spin response of Kerr}

\author{Pierre Vanhove}
\affiliation{Institut de Physique Th\'eorique, CEA, CNRS, Universit\'e Paris-Saclay, F-91191 Gif-sur-Yvette, France}
\affiliation{Laboratoire d’Annecy-le-Vieux de Physique Th\'eorique (LAPTh), CNRS and Universit\'e Savoie Mont-Blanc, Annecy, 74940, France}

\date{\today}

\begin{abstract}
The radial Teukolsky equation is invariant under exchange of the outer and inner
horizons, although the physical boundary condition is not. We derive an exact
monodromy relation between the corresponding Mano--Suzuki--Takasugi (MST)
near-zone responses. At leading order in frequency and to all orders in spin,
this relation yields, for every multipole, an expression for the leading
horizon-exchange-odd response. For gravitational perturbations, it determines
the exchange-odd contribution to quadrupolar absorption and organizes the
exchange-even dynamical response into a digamma function whose expansion
accounts for the odd- and even-zeta structures found in Kerr-Compton
calculations. The exact relation, rather than the individual previously known
small-spin coefficients, is the central result.
\end{abstract}

\maketitle

\paragraph*{Introduction.}---The mass of a Kerr black hole can be written in
terms of its irreducible mass $m_{\rm ir}$ and angular momentum $J$
as~\cite{Christodoulou:1970wf,Christodoulou:1972kt}
\begin{equation}
 M^2=m_{\rm ir}^2+\frac{J^2}{4m_{\rm ir}^2}.
 \label{eq:CR}
\end{equation}
It is invariant under
\begin{equation}
 m_{\rm ir}\longleftrightarrow \widetilde m_{\rm ir}
       =\frac{|J|}{2m_{\rm ir}},
 \qquad A_+=16\pi m_{\rm ir}^2\longleftrightarrow A_- ,
 \label{eq:exchange}
\end{equation}
which exchanges the outer and inner horizon areas. This algebraic exchange is
not a symmetry of black-hole scattering. The physical solution is selected by
an ingoing boundary condition at the event horizon, whereas the inner horizon
lies behind it. Nevertheless, as we show, the radial perturbation equation is
invariant under the exchange. It therefore maps the physical solution to a
well-defined image solution, and the two associated tidal responses obey an
exact relation.

This comparison has a direct physical use. Black-hole perturbation theory (BHPT)
has produced small-spin Kerr-Compton data below extremality and
continuation-dependent coefficients above
extremality~~\cite{Bautista:2021wfy,Bautista:2022wjf,Bautista:2023sdf,
Bautista:2024pn,BjerrumBohr:2026kc}. BHPT data through sixth order in spin are
already used in the aligned-spin analysis of Ref.~\cite{Bautista:2024pn}, and
recently reassembled in Ref.~\cite{BjerrumBohr:2026kc}. We give an exact
relation that explains how the two horizon-selected responses are related,
together with its general-multipole consequence in the long-wavelength regime
with the black-hole spin left unconstrained. The far-zone propagation is common
to the two solutions and all dependence on which horizon condition is imposed is
carried by the near-zone response. In the low-frequency quadrupole, the
exchange-odd term changes the scattering probability and gives a definite
interference contribution to physical absorption. The exchange-even response
contains the conservative phase and contact information, but need not be purely
conservative: it can also contain exchange-even absorption, in particular in the
superradiant regime.

\paragraph*{Horizon exchange and response.}---For Kerr, write
$r_\pm=M(1\pm\kappa)$, with $a = J/M$ the specific angular momentum, $q=a/M$ the
dimensionless spin and $\kappa=\sqrt{1-q^2}$. Exchanging the horizons is
$\kappa\mapsto-\kappa$. The radial Teukolsky equation \cite{Teukolsky:1973ha},
\begin{multline}
\Delta^{-s}\frac{d}{dr}\left(\Delta^{s+1}\frac{dR}{dr}\right)\\
+\left[\frac{K^2-2is(r-M)K}{\Delta}+4is\omega r-\lambda\right]R=0,
\end{multline}
where $m$ is the azimuthal mode number and
$K=(r^2+a^2)\omega-am$, depends on the horizons only through $M$ and $a$ and is
consequently invariant. The local horizon exponents are governed by
\begin{equation}
\epsilon_\pm=\frac{\omega-m\Omega_\pm}{4\pi T_\pm}
=\frac{\epsilon\pm\tau}{2},\quad
\epsilon=2M\omega,\quad \tau=\frac{\epsilon-mq}{\kappa},
\label{eq:exponents}
\end{equation}
with $\Omega_\pm=a/(2Mr_\pm)$ and $T_\pm=(r_\pm-r_\mp)/(8\pi Mr_\pm)$. The
angular velocity of the event horizon is $\Omega_H:=\Omega_+$ and $T_H:=
T_+$ is its Hawking temperature. Thus $\tau$ changes sign under the exchange.
The physical solution $R_+$ is ingoing at $r_+$ and its image $R_-^{(\eta)}$ is
selected by the corresponding ingoing local exponent at $r_-$. At large radius
write each solution in the MST Floquet basis
$R_C^\nu,R_C^{-\nu-1}$~\cite{Mano:1996vt,Mano:1996gn,Sasaki:2003xr}:
\begin{equation}
\begin{aligned}
 R_+&=A_+R_C^\nu+B_+R_C^{-\nu-1},\quad &\Kc_+&:=\frac{B_+}{A_+},\\
 R_-^{(\eta)}&=A_-^{(\eta)}R_C^\nu+B_-^{(\eta)}R_C^{-\nu-1}, \quad
 &\Kc_-^{(\eta)}&:=\frac{B_-^{(\eta)}}{A_-^{(\eta)}}.
\end{aligned}
 \label{eq:K}
\end{equation}
The ratio $\Kc_+$ is the MST near-zone (tidal-response) coefficient selected by
the physical event-horizon boundary condition, while $\Kc_-^{(\eta)}$ is the
corresponding coefficient for the auxiliary inner-horizon image. Here and below,
``near zone'' and ``far zone'' refer operationally to the two factors in the MST
factorization of the partial-wave $S$ matrix, not to a sharp division of the
radial coordinate. The near-zone factor is the response-dependent factor
containing $\Kc$, which carries the information about the horizon boundary
condition. The far-zone factor is the remaining propagation and asymptotic
normalization factor, common to the two horizon-selected solutions. In a
low-frequency matched-asymptotic interpretation, these factors encode,
respectively, the horizon-sensitive response and propagation from the matching
region to infinity. This terminology follows the factorization used in
Ref.~\cite{Bautista:2023sdf_nzff}.

\paragraph*{Exact exchange law.}---Our main result is the following.
For integer spin weight $s$, generic frequency ($2\nu\notin\mathbb Z$ and
$\epsilon_\pm\ne0$), and an image solution continued from $r_-$ to infinity
along a path labelled by $\eta=\pm1$, the Floquet exponent and far-zone
scattering data are exchange-invariant, while the near-zone responses satisfy
\begin{equation}
\frac{\Kc_-^{(\eta)}}{\Kc_+}
=e^{i\pi\eta(2\nu+1)}
 \frac{\sin\pi(\nu-i\tau)}{\sin\pi(\nu+i\tau)} .
\label{eq:law}
\end{equation}
\begin{figure}[t]
\includegraphics[width=\columnwidth]{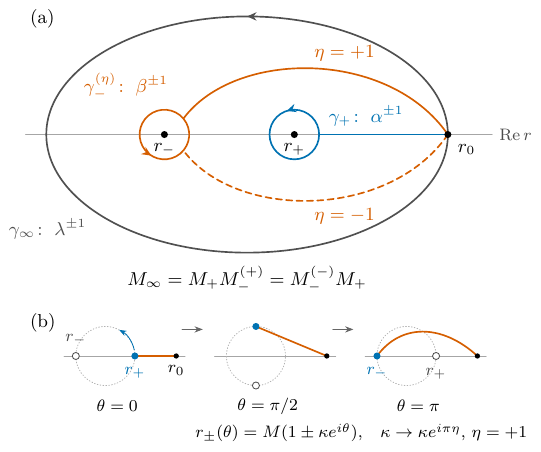}
\caption{Monodromies of the radial equation in the complex $r$ plane, based at
$r_0>r_+$; all loops are counterclockwise. (a)~Eigenvalues and eigenvectors:
$\mathfrak a^{\pm1}$ for $\gamma_+$ (physical and outgoing solutions); $\mathfrak b^{\pm1}$
for $\gamma_-^{(\eta)}$ (image $R_-^{(\eta)}$), whose tail passes $r_+$ above for
$\eta=+1$ and below for $\eta=-1$; $\lambda^{\pm1}$ for $\gamma_\infty$ (Floquet
solutions). (b)~The exchange $\kappa\to\kappa e^{i\pi}$ ($\eta=+1$) turns the
event horizon (filled dot) through ${\rm Im}\,r>0$ into $r_-$ and drags the
continuation path above $r_+$.}
\label{fig:monodromy}
\end{figure}
The two values of $\eta$ specify the two homotopy classes of continuation
around the outer horizon (Fig.~\ref{fig:monodromy}). They are path labels, not
black-hole parameters. They correspond to the two analytic continuations of
the inner-horizon solution and should not be confused with the $\alpha$
bookkeeping labels used for particular contact-term contributions in the
Kerr-Compton literature. The ratio is independent of the normalization of
either radial solution or of the common Floquet basis. Equation~\eqref{eq:law}
is exact in frequency and follows from the $2\times2$ monodromy algebra, not
from a low-frequency expansion or an explicit connection formula. 

To see the origin of the relation, let $M_\pm$ denote the monodromies around
the two horizons and $M_\infty$ the monodromy enclosing both, based at a point
$r_0>r_+$ [Fig.~\ref{fig:monodromy}(a)]. In the Floquet
basis, $M_\infty=\mathrm{diag}(\lambda,\lambda^{-1})$, with
$\lambda=e^{2\pi i\nu}$. The physical solution and its outgoing partner are
eigenvectors of $M_+$ with eigenvalues
$\mathfrak a^{\pm1}$, where $\mathfrak a=e^{2\pi\epsilon_+}$. The image is an
eigenvector of $M_-$ with eigenvalue $\mathfrak b=e^{2\pi\epsilon_-}$. Using
$M_\infty=M_-M_+$ or $M_+M_-$, the trace condition
$\mathrm{tr}\,M_-=\mathfrak b+\mathfrak b^{-1}$ fixes the second eigenvector of $M_+$.
The resulting image eigenvector gives
\begin{equation}
\frac{\Kc_-^{(\eta)}}{\Kc_+}
=-\lambda^\eta\frac{\lambda\mathfrak a-\mathfrak b}{\lambda\mathfrak b-\mathfrak a},
\label{eq:monodromy}
\end{equation}
which is Eq.~\eqref{eq:law}, since $\mathfrak a/\mathfrak b=e^{2\pi\tau}$. The full
proof, including conventions and path ordering, is provided in
Supplemental Sec.~\ref{supp:S1}~\cite{SM}.

At zero frequency, $\nu=\ell$ and the exchange law becomes trivial. This is
consistent with the polynomial truncation of the static perturbation problem
and with the vanishing of the four-dimensional static Kerr Love numbers
established in Refs.~\cite{LeTiec:2020spy,Chia:2020yla,
Charalambous:2021mea,Charalambous:2021kcz,Hui:2021vcv}. The triviality of the
monodromy relation in this limit is a consistency observation. It is not, by
itself, a matching calculation or a proof that the corresponding effective
worldline Wilson coefficients vanish. In the EFT description, static Love
numbers are defined by matching the low-frequency scattering amplitude after
separating universal long-range contributions. Recent amplitude-based matching
calculations confirm the vanishing of the static Schwarzschild Love numbers
on shell~\cite{Correia:2026gravitational,Brunello:2026compton}; these
Schwarzschild results provide a complementary EFT perspective, rather than an
independent derivation of the rotating Kerr result. The non-trivial dynamics
and the running of the dynamical tidal response are controlled by
$\delta=\nu-\ell=O(\epsilon^2)$~\cite{Saketh:2023bul,Perry:2023wmm}.
The MST expression for the partial-wave scattering matrix can be written
\begin{equation}
S_{\ell m}(\Kc)=F_{\ell m}\times
\frac{1+i e^{i\pi\nu}\Kc}
     {1-i e^{-i\pi\nu}\rho\Kc},\quad
\rho=\frac{\sin\pi(\nu-s+i\epsilon)}
           {\sin\pi(\nu+s-i\epsilon)},
\label{eq:Smatrix}
\end{equation}
where $F_{\ell m}$ is the horizon-independent far-zone propagation and
asymptotic-normalization factor. The rational factor is the near-zone response
factor and depends either on $\Kc_+$ or on $\Kc_-^{(\eta)}$. The label
``far zone'' here refers to this factor in the MST partial-wave
factorization, not to the complete
point-particle Compton amplitude or the complete covariant contact polynomial. 

\paragraph*{Universal fixed-$a\omega$ response at arbitrary multipole.}---The exact
law~\eqref{eq:law} has a particularly simple fixed-$a\omega$ limit. Take
$\epsilon=2M\omega\to0$ at fixed $c=a\omega$, and write $\delta=\nu-\ell$ and
$u=i\tau+m$ where $\tau$ is defined in~\eqref{eq:exponents}, both of order
$\epsilon^2$. Define the near-zone factor by $\mathcal N^{\rm NZ}_{\ell
m}(\eta)=1+H^{\rm even}_{\ell m} +\eta H^{\rm odd}_{\ell m}$, so that $H^{\rm
odd}_{\ell m}$ is one half of the difference between the two continuation
choices. The exact exchange law then fixes its leading term for every integer
$\ell$ and $|m|\leq\ell$:
\begin{equation}
H^{\rm odd}_{\ell m}=-\frac{(-1)^m}{2}(4ia\omega)^{2\ell+1}
\mathcal N_{\ell m}\,\epsilon\,R_{\ell m}(a\omega)+O(\epsilon^2),
\label{eq:general-odd}
\end{equation}
where
\begin{equation}
\mathcal N_{\ell m}=\frac{(\ell+m)!(\ell-m)!
[(\ell-s)!(\ell+s)!]^2}{[(2\ell)!(2\ell+1)!]^2},
\end{equation}
and $R_{\ell m}(c):=\lim_{\epsilon\to0}\mathcal X_{-\nu-1}/\mathcal X_\nu$,
where the limit is taken at fixed $c=a\omega$ (Supplemental Sec.~\ref{supp:S2}),
with $R_{\ell m}(0)=1$. Equation~\eqref{eq:general-odd} is the
all-orders-in-$a\omega$ result at leading order in $M\omega$, not a
finite-frequency formula for the complete scattering amplitude.
Its important content is that the dependence on the anomalous dimension $\delta$
cancels between the monodromy relation and the MST near-zone factor. The
remaining function $R_{\ell m}$ is determined by the limiting MST recurrence
(Supplemental Sec.~\ref{supp:S2}). For gravitational perturbations
($s=-2$), its quadrupolar expansion begins
\begin{equation}
R_{2m}=1+\frac{10m}{9}a\omega+
\frac{5(475m^2+261)}{3969}(a\omega)^2+O((a\omega)^3).
\label{eq:R2m-main}
\end{equation}
Equation~\eqref{eq:general-odd} is the main all-multipole result of the paper in
the fixed-$a\omega$, leading-small-$M\omega$ regime: it gives the leading
exchange-odd response to all orders in $a\omega$, with the entire multipole
dependence encoded in the limiting MST series ratio $R_{\ell m}$. The
quadrupolar Bose-factor expression below and the polar closed-form
amplitude~\eqref{eq:closed-odd} are special projections of it.

Thus the far-zone contribution cancels when the two horizon-selected amplitudes
are subtracted, and the difference is sourced by the change in the near-zone
response. Define $S^-_\eta=[S(\Kc_+)-S(\Kc_-^{(\eta)})]/2$ and expanding the
near-zone factor to first order in $\Kc_+$ gives
\begin{equation}
S^-_\eta =\frac{i}{2}F E\left(\Kc_+-\Kc_-^{(\eta)}\right)
       +O(\Kc_+^2),\quad
E=e^{i\pi\nu}+e^{-i\pi\nu}\rho.
\label{eq:Sodd}
\end{equation}
Here $E$ is the linear coefficient of the near-zone factor in the MST $S$
matrix. The far-zone factor $F$ is needed to normalize the full scattering
amplitude, but it carries no horizon-exchange odd information as this is provided by
the factor $E(\Kc_+-\Kc_-^{(\eta)})$ which is the leading change
in the scattering amplitude caused by replacing the event- horizon condition by
its inner-horizon image. Near $\nu=\ell$ both $E$ and $1-\mathcal
K_-^{(\eta)}/\Kc_+$ vanish linearly in $\delta=\nu-\ell$, while $\mathcal
K_+$ has a double pole. Therefore, the $E(\Kc_+-\Kc_-^{(\eta)})$ starts at
$O(\epsilon^{2\ell+2})$. This is the lifting of the static polynomial-truncation
degeneracy. This linearization is used only for the low-frequency quadrupolar
extraction below. The all-orders-in-$a\omega$ result~\eqref{eq:general-odd}
in the leading-small-$M\omega$ regime is instead obtained from the full MST
near-zone factor and the exact exchange law.

For gravitational quadrupole perturbations ($s=-2$, $\ell=2$), the leading term
can be evaluated analytically. With $\varpi=m\Omega_H/(2\pi T_H)=mq/\kappa$, for
the continuation $\eta=+1$, and after removing the common long-range phase,  we
obtain
\begin{equation}
S^-\big|_{\epsilon^6}=-\frac{\pi\kappa^5}{450}
\frac{\varpi(1+\varpi^2)(4+\varpi^2)}
     {e^{2\pi\varpi}-1},
\label{eq:odd}
\end{equation}
which is real at this order.
The horizon Bose factor makes the thermal weighting of the horizon response
explicit.  In the unit-flux partial-wave convention (see the Supplemental
Sec.~\ref{supp:S8}), we use the parity-resolved reflection coefficient
$\mathcal S_P$ for flux balance. The helicity-preserving coefficient $S$
used in the Compton amplitude differs by the helicity-reversing channel.
Their conversion is exchange-even, so it leaves the leading exchange-odd
term below unchanged. The interference of the exchange-odd term with the
exchange-even amplitude gives
\begin{multline}
 \Delta_{\rm odd}\mathcal P_H
 =-2\operatorname{Re}\!\left[(S^{\rm even})^*S^-\right]\\
 =\frac{\pi\kappa^5}{225}
 \frac{\varpi(1+\varpi^2)(4+\varpi^2)}
      {e^{2\pi\varpi}-1}\,\epsilon^6
 +O(\epsilon^7),
 \label{eq:odd-absorption}
\end{multline}
where $S^{\rm even}=1+O(\epsilon)$ after the common phase is removed. Here
$\Delta_{\rm odd}\mathcal P_H$ denotes the term linear in the exchange-odd
amplitude in the expansion of $1-|S_+|^2$. This is an interference contribution,
not a separately measurable absorption probability. The full absorption
probability also contains the exchange-even contribution and, beyond the
helicity-preserving normalization, the helicity-reversing channel described in
Supplemental Sec.~\ref{supp:S8}. For $\eta=-1$ the Bose factor $\bar
n=1/(e^{2\pi\varpi}-1)$ is replaced by $\bar n+1$, and the difference,
$(\pi/225)\,mq[1+(m^2-1)q^2][4+(m^2-4)q^2]\,\epsilon^6$, is polynomial in the
spin. In the Schwarzschild limit this exchange-odd contribution is
$2\epsilon^6/225$, exactly one half of Page's quadrupolar absorption probability
$4\epsilon^6/225$~\cite{Page:1976df}, with the other half provided by the
exchange-even sector. For $m=0$ the exchange-odd share is $\kappa/(1+\kappa)$.
This is a useful quantitative illustration that horizon-exchange parity is not
identical to the conservative/absorptive split. More generally, the
exchange-even sector can also change the modulus of $S_{\ell m}$. In particular,
the leading superradiant term $2q(1+3q^2)\epsilon^5/225$ of $|S_{22}|$ is
exchange-even~\cite{Saketh:2023bul}. The exchange parity is therefore a
classification of the response under the auxiliary horizon map, not a synonym
for the real/imaginary or conservative/ absorptive decomposition at all orders.

The physical interpretation is constrained by the standard flux balance.
For the event-horizon solution, the Teukolsky--Starobinsky identities convert
a conserved cross-spin radial Wronskian into
$\mathcal F_{\rm in}^{\infty}-\mathcal F_{\rm out}^{\infty}
=\mathcal F_H$, with $\mathcal F_H$ proportional to
$(\omega-m\Omega_H)$ times a non-negative horizon-amplitude norm
in a flux-normalized convention~\cite{TeukolskyPress:1974}. Thus the horizon
energy flux changes sign at the superradiant threshold. This does not make
$S^-$ a difference of two horizon fluxes because the inner-horizon image is not a
second solution in the physical exterior scattering problem, and the flux is
quadratic in the perturbation. Rather, $S^-$ measures the change in the
scattering amplitude induced by changing the local horizon condition. Its
real leading term changes the physical scattering probability and hence
contributes to the flux balance. The next term in the low-frequency
expansion and the unit-flux normalization are given in Supplemental
Secs.~\ref{supp:S6} and~\ref{supp:S8}.

\paragraph*{Conservative response and zeta values.}---The exchange-even
projection contains a contribution from the far zone as well as the
near-zone response. Combining them, the exchange-even quadrupolar
coefficient at order $\epsilon^6$ is
\begin{equation}
A_6^++B_6\ln\epsilon
=\frac{i}{450}\Pi_m
 \left[\ln(\epsilon\kappa)+\mathrm{Re}\,\psi(1+i\varpi)\right]
+P_5(q),
\label{eq:even}
\end{equation}
where $P_5(q)$ denotes a degree-five polynomial and
\begin{align}
\Pi_m&=\kappa^5\varpi(1+\varpi^2)(4+\varpi^2)\nonumber\\
&=mq[1+(m^2-1)q^2][4+(m^2-4)q^2].
\end{align}
The digamma and logarithmic terms are derived analytically from the exchange law
and the MST representation. The polynomial part is identified
numerically~\cite{Vanhove:horizonexchange:code} with its continuation tested
against independent MST data on the superextremal branch (Supplemental
Sec.~\ref{supp:S4}). The logarithmic coefficient describes the running of the
dynamical Love response, while the digamma term gives its non-polynomial spin
dependence. In this sense the exchange construction does not identify
conservative physics with a formal even projection. It supplies the
horizon-sensitive piece, which must be combined with the invariant far-zone
phase to obtain the conservative response.

The separation into real and imaginary parts of the digamma function explains
the transcendentality pattern:
\begin{align}
\mathrm{Re}\,\psi(1+i\varpi)
 &=-\gamma_E+\sum_{k\ge1}(-1)^{k+1}\zeta(2k+1)\varpi^{2k},\\
\mathrm{Im}\,\psi(1+i\varpi)
 &=\sum_{k\ge1}(-1)^{k+1}\zeta(2k)\varpi^{2k-1}.
\end{align}
Thus the conservative spin coefficients contain odd zeta values, whereas the
absorptive coefficients contain even zeta values. This transcendental pattern
and the relevant low-spin coefficients have antecedents in the BHPT and
Kerr-Compton literature~\cite{Bautista:2024pn,BjerrumBohr:2026kc}. 
The new statement is that the
horizon-exchange law organizes the two structures as projections of the same
near-zone response.

\paragraph*{Connection across extremality.}---For incidence along the spin, only
$m=2$ contributes. The digamma contribution to the quadrupolar conservative
contact coefficient admits the all-orders-in-spin expression
\begin{equation}
\mathcal C(q)=\frac{128}{45}q(1+3q^2)
\left[\ln\kappa+\mathrm{Re}\,\psi\left(1+\frac{2iq}{\kappa}\right)\right],
\label{eq:C}
\end{equation}
recalling that $\kappa=\sqrt{1-q^2}$.
Its low-spin expansion reproduces the BHPT coefficients
of~\cite{Bautista:2024pn}, and the all-orders resummation in~\eqref{eq:C}
is the new result.
Analytically continuing the symmetrized digamma representation across
$q=1$ gives, for $q>1$,
\begin{multline}
\mathcal C_{\rm SE}(q)
={}\frac{128}{45}q(1+3q^2)
\bigg[
\ln\sqrt{q^2-1}\\
\quad+\frac12\left\{
\psi\left(1+\frac{2q}{\sqrt{q^2-1}}\right)
+\psi\left(1-\frac{2q}{\sqrt{q^2-1}}\right)
\right\}
\bigg].
\label{eq:CSE}
\end{multline}
The pole of the second digamma function at argument $-1$ controls the
large-spin limit,
\begin{equation}
\lim_{q\to\infty}\frac{\mathcal C_{\rm SE}(q)}{q^5}
=\frac{64}{15}.
\label{eq:CSElimit}
\end{equation}
This reproduces the corresponding $\alpha$-tagged coefficient of
Ref.~\cite{Bautista:2022wjf}. Here $\alpha$ is the bookkeeping label in that
amplitude decomposition for terms associated with the continued digamma
contribution. Superextremality itself is specified by $q>1$. The poles of
the continued digamma reproduce the corresponding $\alpha$-tagged terms at
fifth and sixth order in spin, while Eq.~\eqref{eq:odd} reproduces the
continuation-odd coefficients. The branch prescription, pole structure, and
large-$q$ derivation are given in Supplemental Sec.~\ref{supp:S4}.

\paragraph*{Seventh order and beyond.}---At fixed $a\omega$, the exchange law also determines
the leading small-$M\omega$ exchange-odd response for general $\ell$. Combining
this result for $\ell=2$ with the spheroidal-harmonic mixing and the leading
$\ell=3$ contribution gives, for incidence along the spin, the first three
terms of
\begin{multline}
\frac{f^{\rm odd}}{f_0}=-\frac{64}{15}\sin^2\!\frac\Theta2(a\omega)^5
\Bigg[1+\frac23a\omega(3+\cos\Theta)\\
+(a\omega)^2\left(\frac{13}{6}+\frac{29}{21}\cos\Theta
+\frac{11}{42}\cos^2\Theta\right)\\
+\frac{(a\omega)^3}{42}(3+\cos\Theta)(23+14\cos\Theta+3\cos^2\Theta)\\
+O((a\omega)^4)\Bigg],
\label{eq:prediction}
\end{multline}
where $\Theta$ is the scattering angle and $f_0$ the spinless amplitude. The
first two terms reproduce the sixth-order result, Eq.~(3.52) of
Ref.~\cite{Bautista:2022wjf} for the superextremal analytic continuation, and
the $a^7$ and $a^8$ coefficients (second and third lines) agree with Eq.~(5.5)
of Ref.~\cite{Bautista:2023sdf_nzff}, which extends this to eighth order in
spin. For this configuration the higher orders are fixed as well. The limiting
amplitude computed independently from the full MST partial-wave sum
($\ell\le19$, 110 digits) is reproduced order by order through $O(a^{13})$, to
relative accuracy $10^{-15}$, by the closed form
\begin{equation}
\frac{f^{\rm odd}}{f_0}=-\frac{64c^4}{3}\,e^{-2cv}\!\int_0^1\!dt\,t^3e^{4ct}
\big(e^{4cv(1-t)}-1\big),
\label{eq:closed-odd}
\end{equation}
with $c=a\omega$ and $v=\sin^2(\Theta/2)$. Its expansion confirms the $a^7$
coefficient and gives the $a^8$ term on the third line of
Eq.~\eqref{eq:prediction} with the next orders listed in Supplemental
Sec.~\ref{supp:S3}. Equation~\eqref{eq:closed-odd} was identified from
high-precision MST data and independently checked against the segment
representation. Its all-order closed form is therefore a strongly validated
reconstruction, rather than a consequence already proved by the monodromy
algebra alone. In the formal $M\to0$ limit at fixed $a$ and $\omega$, the horizon locations
approach $r=\pm ia$, and the exchange-odd parts of the partial-wave $S$-matrix elements are
given exactly by an integral along the segment joining them plus an endpoint
term [Supplemental Eq.~\eqref{eq:segment}], which we checked against MST to
$10^{-16}$. Correspondingly, Eq.~\eqref{eq:closed-odd} is a superposition of
$e^{-2a'\omega v}$, the exponential of Ref.~\cite{Bautista:2022wjf} for spin
$a'$ at this kinematics, with $a'$ ranging over $[-a,a]$ [Supplemental
Eq.~\eqref{eq:superposition}].

\paragraph*{Discussion.}---The exact response law~\eqref{eq:law} is the structural theorem, and
Eq.~\eqref{eq:general-odd} is its principal physical consequence giving a universal
all-orders-in-spin expression for the leading exchange-odd response at every
multipole. Horizon exchange is not a symmetry between two physical Kerr scattering
experiments because the inner-horizon solution is an auxiliary solution of the same
radial equation, selected by the exchanged local boundary condition. The
far-zone propagation is common to the two solutions, so their difference
isolates the change in the near-zone response. At leading quadrupolar order
this difference is real after removal of the common long-range phase and
therefore contributes to the physical scattering probability and flux
balance. It is an interference contribution, not a separate horizon-flux
probability. The exchange-even sector contains the conservative phase/contact
information but can also contain absorption, including the leading
superradiant term. Horizon-exchange parity is therefore not identical to the
conservative/absorptive decomposition.

The digamma response organizes the previously computed subextremal and
superextremal Kerr-Compton coefficients. The superextremal regime is the
analytic continuation $q>1$ with imaginary $\kappa$, this is not represented by
setting a contact-term bookkeeping parameter to a particular value. In
particular, the far-zone point-particle amplitude and its covariant contact
polynomial must be distinguished from the near-zone response constrained by
Eq.~\eqref{eq:law}. The polar closed form above fixes a polar projection only,
not every independent far-zone contact coefficient.
We reproduce the digamma origin of the non-rational spin
dependence~\cite{BjerrumBohr:2026kc}, the $\alpha$ label on a digamma
function~\cite{Bautista:2023sdf_nzff}, and the monodromy approach to Kerr
scattering coefficients~\cite{Castro:2013scatt,daCunha:2015kerr}.
The new results derived here are the exact law and its general-multipole
consequence~\eqref{eq:general-odd} in the fixed-$a\omega$,
leading-small-$M\omega$ regime, together with the exchange-parity
organization it induces. The exact law, the general-multipole formula, and the leading quadrupolar
exchange-odd response are analytic results. The polynomial part of the conservative coefficient is a numerical identification
supported by high-precision continuation. Extending the closed-form analysis to
higher multipoles, deriving the remaining polynomial terms analytically, and
deriving Eq.~\eqref{eq:closed-odd} by summing the segment representation over
multipoles are important next steps.

\paragraph*{Acknowledgments.}---We thank F.~Bautista for feedback and discussions that led to improvements in
the manuscript. This work has been supported by the Agence Nationale de la
Recherche (ANR) under grant Observables (ANR-24-CE31-7996). We acknowledge the
use of Claude, an AI model developed by Anthropic, to proofread the text and to
write and run the Python code for the numerical checks quoted in the
Supplemental Material. The author takes full responsibility for the content of
the paper.

\textit{Code availability.}---The scripts and data reproducing 
the numerical checks of this paper are available at~\cite{Vanhove:horizonexchange:code}.

\clearpage
\onecolumngrid
\section*{Supplemental Material}
\setcounter{equation}{0}
\renewcommand{\theequation}{S\arabic{equation}}
\setcounter{table}{0}
\renewcommand{\thetable}{S\arabic{table}}

The Supplement establishes the monodromy exchange law, derives its
fixed-$a\omega$, small-$M\omega$ and low-frequency consequences, and documents
the numerical reconstruction and flux normalization. Each equation is
introduced where its definitions and derivation are discussed.
All scripts and data are available at~\cite{Vanhove:horizonexchange:code}.

\suppsection{supp:S1}{The exchange law: setting and proof.}
\emph{Setting.} For integer $s$ the radial Teukolsky equation has regular
singular points at $r=r_\pm$, with exponents
$\{-s-i\epsilon_\pm,\,i\epsilon_\pm\}$, and an irregular singular point at
$r=\infty$, as in the monodromy approach to Kerr
scattering~\cite{Castro:2013scatt,daCunha:2015kerr}. Fix a base point $r_0>r_+$
on the real axis. The monodromy $M_\infty$ along a large circle through $r_0$
has eigenvalues $\lambda^{\pm1}$, $\lambda=e^{2\pi i\nu}$, with eigenfunctions
the Floquet solutions $R_C^{\nu}$, $R_C^{-\nu-1}$; for $2\nu\notin\mathbb Z$
these are unique up to their normalizations, and in this basis $M_\infty={\rm
diag}(\lambda,\lambda^{-1})$. Let $M_+$ be the monodromy along a loop from $r_0$
around $r_+$ alone. The physical solution $R_+$, defined near $r_+$ by its
exponent $-s-i\epsilon_+$, is the eigenvector of $M_+$ with eigenvalue
$\mathfrak a=e^{2\pi\epsilon_+}$, and the outgoing solution at $r_+$ is the
eigenvector with $\mathfrak a^{-1}$; for integer $s$, $\det M_+=1$, and
$\mathfrak a\neq\mathfrak a^{-1}$ for $\epsilon_+\neq0$. The image $R_-$ is defined near
$r_-$ by the exponent $-s-i\epsilon_-$ and continued to $r_0$ along a path
passing $r_+$ on one side. The corresponding loops around $r_-$, with tails
passing $r_+$ above ($\eta=+1$) or below ($\eta=-1$), give two monodromies
$M_-^{(\eta)}$ (Fig.~\ref{fig:monodromy}). With the monodromy of a path $\gamma$
followed by $\gamma'$ equal to $M_{\gamma'}M_\gamma$, so that monodromies act on
the coefficient vectors of solutions in the Floquet basis,
$M_\infty=M_-^{(-)}M_+=M_+M_-^{(+)}$; hence
$M_-^{(+)}=M_+^{-1}M_-^{(-)}M_+$, and the two images differ by the action of
$M_+$. $R_-^{(\eta)}$ is the eigenvector of $M_-^{(\eta)}$ with eigenvalue
$\mathfrak b=e^{2\pi\epsilon_-}$.

\emph{Convention independence.} Writing each solution as $R\propto R_C^\nu+\Kc
R_C^{-\nu-1}$, a rescaling $R_C^\nu\to c_1R_C^\nu$, $R_C^{-\nu-1}\to
c_2R_C^{-\nu-1}$ maps $\Kc\to(c_1/c_2)\Kc$ for every solution, so $\Kc_-^{(\eta)}/\Kc_+$
is invariant. The normalizations of $R_\pm$ do not enter $\Kc$. The relabelling
$\nu\to-\nu-1$ exchanges the two Floquet solutions, maps $\Kc\to1/\Kc$ and
$\lambda\to\lambda^{-1}$, and maps the right-hand side of Eq.~\eqref{eq:law} to
its inverse. Equation~\eqref{eq:law} is therefore a statement about the pair
$(R_+,R_-^{(\eta)})$ alone.

\emph{Proof.} With $M_+=P\,{\rm diag}(\mathfrak a,\mathfrak a^{-1})P^{-1}$,
$P=\begin{pmatrix}1&1\\ \Kc_+&Y\end{pmatrix}$, the condition that
$M_-^{(-)}=M_\infty M_+^{-1}$ have eigenvalues $\mathfrak b^{\pm1}$ (equivalently
${\rm tr}\,M_-^{(-)}=\mathfrak b+\mathfrak b^{-1}$) gives
\begin{equation}
\frac{Y}{\Kc_+}=\frac{(\lambda\mathfrak a-\mathfrak b)(\lambda\mathfrak a\mathfrak b-1)}{(\lambda-\mathfrak a\mathfrak b)(\lambda\mathfrak b-\mathfrak a)},
\end{equation}
and the $\mathfrak b$-eigenvector $(1,\Kc_-^{(\eta)})$ has
$\Kc_-^{(\eta)}/\Kc_+=-\lambda^{-1}(\lambda\mathfrak a-\mathfrak b)/(\lambda\mathfrak b-\mathfrak a)$. For
$M_-^{(+)}=M_+^{-1}M_\infty$ one finds
$-\lambda(\lambda\mathfrak a-\mathfrak b)/(\lambda\mathfrak b-\mathfrak a)$; the two differ by
$\lambda^2$. With $\mathfrak a=e^{2\pi\epsilon_+}$, $\mathfrak b=e^{2\pi\epsilon_-}$,
$\mathfrak a/\mathfrak b=e^{2\pi\tau}$ and $\lambda=e^{2\pi i\nu}$,
$(\lambda\mathfrak a-\mathfrak b)/(\lambda\mathfrak b-\mathfrak a)=\sin\pi(\nu-i\tau)/\sin\pi(\nu+i\tau)$,
which gives Eq.~\eqref{eq:law}. 
Half-integer $s$, for which $\det M_\pm=-1$, is not covered.

\suppsection{supp:S2}{Fixed-$a\omega$, small-$M\omega$ limit and limiting MST recurrence.}
With the $n=0$ terms of the MST series made explicit~\cite{Bautista:2023sdf_nzff},
\begin{equation}
\Kc=(2\epsilon\kappa)^{2\nu+1}\frac{\Gamma(-2\nu-1)\Gamma(-2\nu)\Gamma(\nu+1-i\tau)\Gamma(\nu+1-s-i\epsilon)\Gamma(\nu+1-s+i\epsilon)}{\Gamma(2\nu+1)\Gamma(2\nu+2)\Gamma(-\nu-i\tau)\Gamma(-\nu-s-i\epsilon)\Gamma(-\nu-s+i\epsilon)}\,\frac{\mathcal X_{-\nu-1}}{\mathcal X_\nu},
\end{equation}
where $\mathcal X_{\nu}$ are ratios of MST series, invariant under the exchange.
The function entering Eq.~\eqref{eq:general-odd} is
$R_{\ell m}(c)=\lim_{\epsilon\to0}\mathcal X_{-\nu-1}/\mathcal X_\nu$
at fixed $c=a\omega$, normalized by $R_{\ell m}(0)=1$. For $\epsilon\to0$ at fixed $a\omega$, $2\epsilon\kappa\to4ia\omega$, $\nu=\ell+\delta$ and $i\tau=-m+u$ with $\delta,u=O(\epsilon^2)$. Using $1/\Gamma(-N-z)\simeq(-1)^{N+1}N!\,z$,
\begin{equation}
\Kc^{(\eta)}\simeq(4ia\omega)^{2\ell+1}\,\frac{(\ell+m)!(\ell-m)!\,[(\ell-s)!(\ell+s)!]^2\,(-1)^{\ell-m}\,\epsilon^2}{4\,\delta^2\,[(2\ell)!(2\ell+1)!]^2}\,(u+\eta\,\delta)\,R_{\ell m}(a\omega),
\label{eq:Kclass}
\end{equation}
and $e^{i\pi\nu}+e^{-i\pi\nu}\rho\simeq(-1)^\ell\,2i\delta/\epsilon$. Thus the leading small-$M\omega$ exchange-odd response, for general integer $\ell$, is
\begin{equation}
H^{\rm odd}_{\ell m}=-\frac{(-1)^m}{2}(4ia\omega)^{2\ell+1}
\mathcal N_{\ell m}\epsilon R_{\ell m}(a\omega)+O(\epsilon^2),
\label{eq:Hodd}
\end{equation}
where $\mathcal N_{\ell m}=(\ell+m)!(\ell-m)![(\ell-s)!(\ell+s)!]^2/[(2\ell)!(2\ell+1)!]^2$ and $R_{\ell m}$ is the limiting ratio of MST series at fixed $a\omega$. This is the expression used in the leading-small-$M\omega$ amplitude calculation below.

\emph{Limiting recurrence.}
At $\epsilon\to0$, $c=a\omega$ fixed, the MST coefficients become
\begin{equation}
\begin{aligned}
\alpha_n&\to-2c\,(n{+}\ell{+}1{+}s)^2(n{+}\ell{+}1{-}m)/[(n{+}\ell{+}1)(2n{+}2\ell{+}3)],\cr
\gamma_n&\to2c\,(n{+}\ell{-}s)^2(n{+}\ell{+}m)/[(n{+}\ell)(2n{+}2\ell{-}1)],\cr
\beta_n&\to(n{+}\ell)(n{+}\ell{+}1)-E(c)-c^2-2ms^2c/[(n{+}\ell)(n{+}\ell{+}1)],
\end{aligned}
\end{equation}
with $E(c)$ the spin-weighted spheroidal eigenvalue. For $s=-2$, $\ell=2$ the
negative-$n$ coefficients drop out and
\begin{equation}
R_{2m}^{-1}=\left(\sum_{n\ge0}a_n\right)\left(\sum_{n\ge0}(-1)^n\tfrac{(3-m)_n}{(3+m)_n}a_n\right),
\end{equation}
with $a_1=-\tfrac59(3{+}m)c+\tfrac{5m}{81}(3{+}m)c^2$ and
$a_2=\tfrac{5}{49}(3{+}m)(4{+}m)c^2$. Expanding the resulting series ratio gives, for $s=-2$ and $\ell=2$,
\begin{equation}
R_{2m}=1+\frac{10m}{9}a\omega+
\frac{5(475m^2+261)}{3969}(a\omega)^2+O((a\omega)^3),
\label{eq:R2m}
\end{equation}
which supplies the quadrupolar correction entering Eq.~\eqref{eq:Hodd} and the amplitude in Sec.~\ref{supp:S3}.

\suppsection{supp:S3}{The $\eta$-odd amplitude and higher orders.}
The helicity-preserving amplitude is
\begin{equation}
    f={2\pi\over i\omega}\sum_{\ell m}(e^{2i\delta_{\ell m}}-1)\,S^*_{\ell
m}(\nin)S_{\ell m}(\nout),
\end{equation}
 with $S_{\ell m}={}_{-2}S_{\ell m}$ at
$c=a\omega$~\cite{Dolan:2008kf}. Through $O(a^6)$ only $\ell=2$ contributes to
$f^{\rm odd}$, and with $f_0=M\cos^4(\Theta/2)/\sin^2(\Theta/2)$ (frame phase
restored by the addition theorem)
\begin{equation}
\frac{f^{\rm odd}}{f_0}=-\frac{8}{45}\sin^2\!\frac\Theta2\,(a\omega)^5\,\frac{\sum_m(-1)^m(2{+}m)!\,(2{-}m)!\,R_{2m}\,S^*_{2m}S_{2m}}{\sum_m Y^*_{2m}Y_{2m}},
\end{equation}
$Y_{\ell m}={}_{-2}Y_{\ell m}$. At $O(a^7)$ the numerator acquires
$-\tfrac{(a\omega)^2}{3969}\sum_m(-1)^m(3{+}m)!\,(3{-}m)!\,Y^*_{3m}Y_{3m}$. For
$s=-2$, $m=2$ and $x=\cos\Theta$,
\begin{equation}
{S_{22}(0)S_{22}(\Theta)\over Y_{22}(0)Y_{22}(\Theta)}=1+c\left(\frac{2x}{3}-\frac29\right)
+c^2\left(\frac{11x^2}{42}-\frac{2x}{21}-\frac{173}{2646}\right)+O(c^3)
\end{equation}
and 
\begin{equation}
{Y_{32}(0)Y_{32}(\Theta)\over Y_{22}(0)Y_{22}(\Theta)}=\frac75(3x-2),
\end{equation}
 which
give Eq.~\eqref{eq:prediction} through $O(a^7)$. 

\emph{Independent check and
higher orders.} We computed the polar limiting amplitude directly in the
fixed-$a\omega$, small-$M\omega$ regime:
$\epsilon=10^{-16}$ at fixed $c$, full MST (near and far zone) at 110 digits for
$\ell=2,\dots,19$ and both $\eta$, 28 Chebyshev nodes $|c|\le0.2$, projected onto
the polar harmonics. Through $O(a^{13})$ it agrees, to $10^{-15}$ relative
($3\times10^{-16}$ through $O(a^{10})$), with
\begin{equation}
\frac{f^{(\eta)}}{f_0}=e^{-2cv}\Big[1+(1-\eta)\,\frac{64c^4}{3}\int_0^1\!dt\,t^3e^{4ct}\big(e^{4cv(1-t)}-1\big)\Big],\qquad v=\sin^2\frac\Theta2 .
\label{eq:closedform}
\end{equation}
For $\eta=+1$ this is the exponential $e^{-2cv}$, and the $\eta$-odd part is
Eq.~\eqref{eq:closed-odd}. With $Q(b)=b^3-3b^2+6b-6$ the integral is elementary,
\begin{equation}
\frac{f^{\rm odd}}{f_0}=-\frac{e^{-2cv}}{12}\left[\frac{e^{4c}\,Q\big(4c(1-v)\big)+6e^{4cv}}{(1-v)^4}-e^{4c}\,Q(4c)-6\right].
\label{eq:closed-elem}
\end{equation}
Its expansion,
\begin{equation}
-\frac{64}{15}vc^5+\frac{256}{45}v(v-2)c^6-\frac{128}{315}v(11v^2-40v+40)c^7
+\frac{256}{315}v(v-2)(3v^2-10v+10)c^8+O(c^9),
\end{equation}
is Eq.~\eqref{eq:prediction} with $\cos\Theta=1-2v$. Writing
$f^{\rm odd}/f_0=-\tfrac{64}{15}v\,c^5\sum_{k\ge0}b_k(x)\,c^k$, with
$b_0,\dots,b_3$ read off from the bracket of Eq.~\eqref{eq:prediction}, the
orders tested through $O(a^{13})$ are completed by
\begin{equation}
\begin{aligned}
b_4&=\tfrac{1}{1512}\,(1463+1828x+938x^2+228x^3+23x^4),\\
b_5&=\tfrac{1}{1890}\,(3+x)(295+362x+188x^2+46x^3+5x^4),\\
b_6&=\tfrac{1}{166320}\,(32081+60138x+49775x^2+22860x^3+6175x^4+938x^5+65x^6),\\
b_7&=\tfrac{1}{498960}\,(3+x)(11525+21414x+17743x^2+8164x^3+2227x^4+342x^5+25x^6),\\
b_8&=\tfrac{1}{51891840}\,(1146727+2878792x+3291140x^2+2212216x^3+956010x^4+272376x^5\\
&\qquad\qquad\qquad+50308x^6+5576x^7+295x^8).
\end{aligned}
\label{eq:bk}
\end{equation}
The odd-order coefficients contain the factor $3+x$: at $x=-3$ ($v=2$)
Eq.~\eqref{eq:closed-odd} is odd in $c$. This is a statement about the polar projection of the exchange-odd
amplitude only. It does not determine the complete covariant far-zone contact
polynomial: independent contact structures can vanish in the polar limit.
This is consistent with the polar simplification in Ref.~\cite{Cangemi:2023},
where the additional contact deformation is not resolved by that kinematic
slice.

\emph{Segment representation.} The expansion at fixed $a$ and $\omega$
starts at first order in $M$. Expanding the radial operator of the main text as
$T_M=T_0+MT_1+O(M^2)$ gives, for $s=-2$,
\begin{equation}
T_1=-2r\,\partial_r^2+2\,\partial_r-\frac{4iK}{\Delta_0}+\frac{2r\,(K^2+4irK)}{\Delta_0^2},
\qquad \Delta_0=r^2+a^2,\quad K=\omega\Delta_0-am .
\label{eq:T1}
\end{equation}
At $M=0$ the radial equation is the spheroidal equation. Its solution regular at
both $r=\pm ia$ is $R_0(r)=\Delta_0(r)\,\hat S_\ell(ir/a)$, where $\hat S_\ell(x)$
is $S_{\ell2}$ as a function of $x=\cos\theta$, continued to complex $x$ (it is
entire). In this limit $r_\pm\to\pm\eta\,ia$, so the continuation $\eta$ imposes
the ingoing condition at $r=\eta\,ia$. The local exponents are $\{3,-1\}$ at
$r=ia$ and $\{1,1\}$ at $r=-ia$, and the ingoing ones are shifted by $-iM\omega$.
Writing both solutions as $R_0+MR_1^{(\eta)}$ and integrating the Green identity
of the formally self-adjoint operator $\Delta^sT_M$ from one horizon to the other
gives the $\eta$-odd part of the partial-wave element,
\begin{equation}
e^{2i\delta_{\ell2}}\big|_{\rm odd}=\frac{\epsilon\,G_\ell}{8i\omega^2B^{\rm inc}B^{\rm ref}},\qquad
G_\ell={\rm FP}\!\int_{-ia}^{ia}\frac{R_0\,T_1R_0}{\Delta_0^2}\,dr
+\hat S_\ell(1)\big[(2c-1)\hat S_\ell(1)-4\hat S_\ell'(1)\big],
\label{eq:segment}
\end{equation}
where $B^{\rm inc}$ and $B^{\rm ref}$ are the coefficients of $r^{-1}e^{-i\omega r}$
and $r^3e^{i\omega r}$ in $R_0$. The integrand has a double pole without residue
at $r=-ia$, and FP denotes its finite part. The last term is the boundary term at
the degenerate horizon $r=-ia$; its $2c$ comes from the ingoing exponent, and the
boundary term at $r=ia$ vanishes. On the segment the radial function is the
angular one, $R_0(ia')=(a^2-a'^2)\,\hat S_\ell(-a'/a)$, and the endpoint term
involves only $\hat S_\ell$ at $\theta=0$. Equation~\eqref{eq:segment} reproduces
the MST values of $e^{2i\delta_{\ell2}}|_{\rm odd}$ computed above, for
$\ell=2,\dots,6$ at two values of $c$, to $4\times10^{-16}$, the size of the
finite-$\epsilon$ systematic. It includes the terms nonlinear in $\Kc$ dropped in
Eq.~\eqref{eq:Sodd}, which are not small here: since $u/\delta=O(c^{-2})$ in
Eq.~\eqref{eq:Kclass}, the exchange-even part of $\Kc$ is enhanced, and these
terms change the $\ell=2$ element already at relative order $c^3$.
Equation~\eqref{eq:closed-odd} can be rewritten as
\begin{equation}
\frac{f^{\rm odd}}{f_0}=-\frac{64c^4}{3}\int_0^1\!dt\,t^3e^{4ct}\big(e^{-2(2t-1)cv}-e^{-2cv}\big),
\label{eq:superposition}
\end{equation}
a superposition of the exponential $e^{-2a'\omega v}$ for spin
$a'=(2t-1)a\in[-a,a]$, which suggests identifying $t$ with the position $ia'$ on
the segment. Likewise
$f^{(-)}/f_0=e^{2cv}/(1-v)^4+e^{2c(2-v)}\big[Q\big(4c(1-v)\big)/(1-v)^4-Q(4c)\big]/6$
contains the exponential for spin $-a$, matching the endpoint $r=-ia$. Summing
Eq.~\eqref{eq:segment} over multipoles would derive Eq.~\eqref{eq:closedform}; we
have not carried this out.

\suppsection{supp:S4}{Low-frequency expansion and numerical reconstruction.}
 \paragraph*{Continuation of the all-orders digamma expression.}
For subextremal real $q$ we have $\kappa=\sqrt{1-q^2}>0$. The real part in
Eq.~\eqref{eq:C} can be written as
\begin{equation}
\mathrm{Re}\,\psi(1+ix)
=\frac12\left[\psi(1+ix)+\psi(1-ix)\right],
\qquad x=\frac{2q}{\kappa}\in\mathbb R.
\label{eq:Re-digamma-sym}
\end{equation}
The right-hand side defines the meromorphic representative to be continued
in complex $q$, away from the digamma poles. Continuing around the branch
point at $q=1$ gives $\kappa=\pm i\lambda$, where
$\lambda=\sqrt{q^2-1}$ on the superextremal real axis. The two digamma
terms are exchanged when the sign of $\kappa$ is reversed, whereas
$\log\kappa=\log\lambda\pm i\pi/2$. Taking the real part on this axis
therefore gives Eq.~\eqref{eq:CSE}. The continuation is understood along
paths in the complex $q$-plane that avoid the poles. On the real
superextremal axis these occur when
\begin{equation}
1-\frac{2q}{\sqrt{q^2-1}}=-k,
\qquad k=2,3,4,\ldots,
\end{equation}
or, equivalently,
\begin{equation}
q=\frac{N}{\sqrt{N^2-4}},
\qquad N=3,4,5,\ldots.
\label{eq:CSE-poles}
\end{equation}

For completeness, the large-$q$ behavior follows from
\begin{equation}
1-\frac{2q}{\sqrt{q^2-1}}
=-1-\frac{1}{q^2}+O(q^{-4}).
\end{equation}
Using the digamma expansion
$\psi(-1+\delta)=-1/\delta+O(1)$ as $\delta\to0$, we obtain
\begin{equation}
\psi\left(1-\frac{2q}{\sqrt{q^2-1}}\right)
=q^2+O(1).
\end{equation}
The other digamma term and the logarithm grow only logarithmically.
Consequently the square bracket in Eq.~\eqref{eq:CSE} is
$q^2/2+O(\ln q)$, which yields Eq.~\eqref{eq:CSElimit}.
This is the $\alpha$-tagged $q^5$ coefficient identified in
Ref.~\cite{Bautista:2022wjf}. The label $\alpha$ there is an amplitude-
decomposition bookkeeping label, not a Kerr parameter and not a label
for either side of extremality. In particular, it should not be confused
with the physical condition $q>1$ or with the path label used for the
continuation.

The physical solution $S_{2m}$ was computed from the MST continued fraction at
90-digit precision, using 12 angular nodes at each of $r=2\times10^{-5}$ and
$10^{-5}$ in $\epsilon$, and 40 ($m=2$) or 24 (otherwise) nodes on $|q|=0.3$.
Both $\kappa$ and $-\kappa$ were evaluated, with
$\log(-\kappa)=\log\kappa+i\pi$; the spheroidal eigenvalue was computed at
complex $c=q\epsilon/2$. The fitted coefficient at order $\epsilon^6$ is
\begin{equation}
A_6+B_6\ln\epsilon=\frac{i}{450}\Pi_m
 [\ln(\epsilon\kappa)+\psi(1+i\varpi)]
-\frac{\kappa^5}{900}(1+\varpi^2)(4+\varpi^2)+p^{(m)}_5(q),
\label{eq:N6}
\end{equation}
where $p^{(m)}_5(q)$ is the degree-five polynomial identified numerically;
its exchange-odd and exchange-even parts are Eqs.~\eqref{eq:odd} and
\eqref{eq:even}.

The logarithmic coefficient is extracted by a joint two-radius fit of the
Fourier coefficients at orders $n=6,7,8$, which removes leakage from higher
powers times $\ln\epsilon$. A fit to $c_6$ alone determines $B_6$ only to
$5\times10^{-5}$--$3\times10^{-4}$, whereas the joint fit gives
$B_6=i\Pi_m/450$ to $10^{-10}$ for every $m$. The exchange-odd terms have no
logarithm through $\epsilon^7$. Independent free fits recover the coefficients
of $\Pi_m\ln\kappa$ and $\Pi_m\operatorname{Re}\psi$ as $-1/450$ to
$3\times10^{-9}$. After subtracting Eq.~\eqref{eq:N6}, fitted Taylor
coefficients for $6\le p\le20$ are below $2\times10^{-11}$, while retained
coefficients are $O(10^{-5}$--$1)$.

The polynomial determined at $|q|\le0.3$ also predicts the super-extremal
branch. For $m=2$, continuation of Eq.~\eqref{eq:even} agrees with direct MST
to $4\times10^{-12}$ for $1.5\le q\le30$ and to $3\times10^{-10}$ for
$50\le q\le5000$ ($6\times10^{-13}$ at $q=5000$). This bounds an additional
$q^6$ coefficient by $3\times10^{-18}$, compared with $2\times10^{-11}$ from
the direct Taylor fit; higher powers are more strongly constrained. The
$q^5$ coefficient of $p_5^{(2)}$, $-2.3457997302\times10^{-5}i$, agrees to
$10^{-15}$ with the exactly known $\eta=+1$ limiting value divided by 32.
For $q>1$, MST data were taken at fixed $q$ on circles of radius proportional
to $1/q$ (three radii for $q\ge50$).

\suppsection{supp:S5}{Contact polynomial and general kinematics.}
Following Refs.~\cite{Bautista:2022wjf,Bautista:2023sdf}, write
$f=f_0(e^{(2w+k_3-k_2)\cdot a}+P_\xi)$, where
\begin{equation}
P_\xi=\sum_{j=0}^{2}\xi^{j-1}(w\cdot a)^{4-2j}[(w-k_2)\cdot a]^j[(w+k_3)\cdot
a]^j\,r^{(j)}+\xi^{-2}(w\cdot a)^6d_1+\xi^2[(w-k_2)\cdot a]^3[(w+k_3)\cdot
a]^3f_1,
\end{equation}
and
\begin{equation}
r^{(j)}=c_3^{(j)}w\cdot
a+c_4^{(j)}|a|\omega+c_6^{(j)}(2w-k_2+k_3)\cdot a\,w\cdot a+c_8^{(j)}(w\cdot
a)^2+c_9^{(j)}(k_2-k_3)\cdot a\,|a|\omega+c_{10}^{(j)}w\cdot a\,|a|\omega
\end{equation}
(terms vanishing through $O(a^6)$ omitted). In the rest frame, for
all-incoming $k_2=\omega(1,\nin)$, $k_3=-\omega(1,\nout)$ and mostly-minus
metric, $\xi^{-1}=-\sin^2(\Theta/2)$, and 
\begin{equation}
w\cdot a=-\omega\,\vec a\cdot(\nin+\nout+i\,\nin\times\nout)/(1+\nin\cdot\nout).
\end{equation}
The far-zone factor in the partial-wave factorization is not itself the
complete covariant contact polynomial $f^{\rm FZ}$: the latter is obtained by
matching the far-zone partial waves to a covariant amplitude at generic
kinematics. A polar restriction cannot fix every independent coefficient,
since contact structures that vanish in that limit remain unconstrained.
The BHPT side is $\sum_m h_m S^*_{2m}S_{2m}$, normalized as in
Sec.~\ref{supp:S3}; $h_m$ is the pole part of the $\ell=2$ element: 
\begin{equation}
h_m^{\rm even}=-{\rm sgn}(m)\,h_m^{\rm odd},\qquad
h_m^{\rm odd}\propto(-1)^m(2{+}m)!(2{-}m)!R_{2m}.
\end{equation}
At random kinematic points the ratios to the $\alpha$-tagged ansatz are
constant to $10^{-30}$, whereas the opposite helicity fails. A free fit of the
full $\eta$-odd amplitude reproduces all nine $\eta$-odd coefficients with
residual $<10^{-23}$; see Table~\ref{tab:coeffs}.

The two uses of $\alpha$ in the literature must not be conflated. In
Refs.~\cite{Bautista:2022wjf,Bautista:2023sdf_nzff}, $\alpha$ labels
contributions associated with analytically continued digamma structure in a
particular decomposition of the superextremal contact terms. The superextremal
regime itself is defined by $q>1$, not by the value of this label. In
Ref.~\cite{Cangemi:2023}, by contrast, $\alpha$ multiplies an additional
contact-term deformation of a candidate Compton amplitude. There $\alpha_{\rm
C}=0$ is the baseline amplitude, which agrees with the BHPT result through the
orders tested, including the terms labelled by the dissipative sign $\eta$ so
that setting $\alpha_{\rm C}\ne0$ adds a distinct contact deformation. Our
horizon-exchange construction has no independent $\alpha_{\rm C}$ parameter: it
determines the full continued near-zone response, including the digamma-pole
contributions. Removing those poles would define a separate subtraction
prescription, not the $\alpha=0$ member of a horizon-exchange family. The
possible identification of the particular contact expression discussed in the
text with the $\alpha_{\rm C}=1$ term of Ref.~\cite{Cangemi:2023} requires a
term-by-term comparison of the full covariant amplitudes in matched kinematic
and normalization conventions; it is not established by the polar limit or by
the present response relation. No identification of the Bautista bookkeeping
label with $\alpha_{\rm C}$ is made here. In particular, our
horizon-continuation label $\eta$ is a path label in the monodromy problem and
is not either contact-term parameter.

\begin{table}[h]
\begin{ruledtabular}
\begin{tabular}{lcccc}
 & & $j=0$ & $j=1$ & $j=2$\\
\hline
$c_3^{(j)}$ & $a^5$, even & $\tfrac{64}{15}$ & $\tfrac{16}{3}$ & $\tfrac{16}{15}$\\[2pt]
$c_4^{(j)}$ & $a^5$, odd & $\tfrac{64}{15}$ & $\tfrac{16}{5}$ & $\tfrac{4}{15}^{*}$\\[2pt]
$c_6^{(j)}$ & $a^6$, even & $\tfrac{128}{45}$ & $\tfrac{32}{9}$ & $\tfrac{32}{45}$\\[2pt]
$c_8^{(j)}$ & $a^6$, even & $-\tfrac{512}{45}$ & $-\tfrac{160}{9}$ & $-\tfrac{304}{45}$\\[2pt]
$c_9^{(j)}$ & $a^6$, odd & $-\tfrac{128}{45}$ & $-\tfrac{32}{15}$ & $-\tfrac{8}{45}^{*}$\\[2pt]
$c_{10}^{(j)}$ & $a^6$, odd & $-\tfrac{256}{45}$ & $-\tfrac{352}{45}$ & $-\tfrac{32}{15}$\\
\end{tabular}
\end{ruledtabular}
\caption{Contact coefficients of Refs.~\cite{Bautista:2022wjf,Bautista:2023sdf} reproduced by the continued $\ell=2$ response ($\eta$-odd entries divided by $\eta$). Entries without asterisk come from the poles of $\psi(1+i\vp)$ and $\coth\pi\vp$ ($\alpha$-tagged in Ref.~\cite{Bautista:2022wjf}); asterisks mark the $m=0$ term of Eq.~\eqref{eq:odd}. In addition $f_1=-16/45$ (pole). The untagged $\eta$-even pieces, e.g.\ $c_3^{(2)}\supset4/15$, are not pure $\ell=2$ and are not tested.}
\label{tab:coeffs}
\end{table}

\suppsection{supp:S6}{Analytic derivation of the absorptive response.}
Write Eq.~\eqref{eq:Smatrix} as $S(\Kc)=F\,[1+H(\Kc)]$ with $H(\Kc)=iE\Kc+O(\Kc^2)$,
$E\equiv e^{i\pi\nu}+e^{-i\pi\nu}\rho$. Since $F$ and $\nu$ are invariant, the exchange-odd part is
$S^-=\tfrac{i}{2}FE\,(\Kc_+-\Kc_-^{(\eta)})$ up to terms quadratic in $\Kc_+$, which are
$O(\epsilon^9)$ here. For integer $s$ and $\nu=\ell+\delta$:

\begin{enumerate}[label=\textup{(\alph*)},leftmargin=2em]
\item $\rho=\sin\pi(\delta+i\epsilon)/\sin\pi(\delta-i\epsilon)$ and, exactly,
$E=(-1)^\ell e^{\pi\epsilon}\sin2\pi\delta/\sin\pi(\delta-i\epsilon)=(-1)^\ell
e^{\pi\epsilon}(2i\delta/\epsilon)[1-i\delta/\epsilon+O(\epsilon^2)]$;
\item
$\Xi=-e^{2i\pi\eta\delta}\,\sin(\pi\delta-i\pi\tau)/\sin(\pi\delta+i\pi\tau)=1+2i\pi\delta(\eta+\coth\pi\tau)+O(\delta^2)$.
This Taylor expansion is at fixed nonzero $\tau$ and is not uniform as
$\tau\to0$. In the axisymmetric and Schwarzschild limits one must retain
the exact expression for $\Xi$ until it is combined with the
$\tau$-dependent factor in $\Kc_+$, and only then take the limit.
In particular,
$\lim_{\tau\to0}\tau(1+\coth\pi\tau)=1/\pi$; the final expression
\eqref{eq:boltzmann} is therefore regular at $\varpi=0$, with
$\varpi/(e^{2\pi\varpi}-1)\to1/(2\pi)$.
For $s=-2$, $\ell=2$: 
\item \emph{Anomalous dimension.} The $n=0$ equation of the
MST recurrence, $\beta_0+\alpha_0R_1+\gamma_0L_{-1}=0$, has no $O(\epsilon)$
term because the spheroidal eigenvalue
$E(c)=6-\tfrac{4m}{3}c+[\tfrac{10}{189}(9-m^2)-1]c^2+O(c^3)$ cancels the
explicit $mq\epsilon$ terms of $\beta_0$. At $O(\epsilon^2)$,
$\alpha_0R_1=-\tfrac{5}{378}\epsilon^2(9\kappa^2+m^2q^2)$,
$\gamma_0L_{-1}=O(\epsilon^4)$, and the $q^2$ terms cancel against those of
$E(c)$, leaving $\delta=\nu_2\epsilon^2+O(\epsilon^3)$ with $\nu_2=-107/210$ for
all $q$ and $m$. Numerically
$\nu=2-\tfrac{107}{210}\epsilon^2+\tfrac{mq}{12}\epsilon^3+O(\epsilon^4)$;
$\nu_3$ does not enter Eq.~\eqref{eq:boltzmann}: once the powers of $\delta$
cancel, $\delta$ appears only through
$\delta/\epsilon=\nu_2\epsilon+O(\epsilon^2)$.
\item \emph{Response.} From Sec.~\ref{supp:S2},
with 
\begin{equation}
\begin{aligned}
(2\epsilon\kappa)^{2\nu+1}&=32\epsilon^5\kappa^5[1+O(\delta\ln\epsilon)],\cr
\Gamma(-4-2\delta)\Gamma(-5-2\delta)&=-[1+O(\delta)]/(11520\,\delta^2),\cr
\Gamma(3-i\tau)/\Gamma(-2-i\tau)&=-i\tau(1+\tau^2)(4+\tau^2),\cr
\Gamma(5-i\epsilon)\Gamma(5+i\epsilon)&=576[1+O(\epsilon^2)],\cr
1/[\Gamma(-\delta-i\epsilon)\Gamma(-\delta+i\epsilon)]&=\epsilon^2[1+O(\delta,\epsilon^2)].
\end{aligned}
\end{equation}
In the series ratio,
$a_1=-\gamma_1/\beta_1=\tfrac{5i}{18}\epsilon\kappa(3-i\tau)$, while
$a_{-1}=O(\epsilon^3)$ because
$\alpha_{-1}\propto(\delta+i\epsilon)(\delta-i\epsilon)$; the $n=1$ terms of
$\mathcal X_\nu$ and of $\mathcal X_{-\nu-1}$ give $\mathcal X_{-\nu-1}/\mathcal
X_\nu=1-\tfrac{5}{9}\epsilon\kappa\tau+O(\epsilon^2)=1+\tfrac{5mq}{9}\epsilon+O(\epsilon^2)$.
Altogether
\begin{equation}
\delta^2\Kc_+=\frac{i}{1800}\,\epsilon^7\kappa^5\,\tau(1+\tau^2)(4+\tau^2)\,\Big[1+\frac{5mq}{9}\,\epsilon+O(\epsilon^2)\Big].
\end{equation}
\item \emph{Far zone.} At $O(\epsilon)$, $F\,e^{-2i\epsilon\ln\epsilon}$ does not
depend on $q$; at $q=0$ it receives $2i\epsilon\ln2$ from $2^{2i\epsilon}$,
$i(2\gamma_E-\tfrac{25}{12})\epsilon$ from
$\Gamma(1-i\epsilon)/\Gamma(5+i\epsilon)$ and $-i\epsilon$ from the $n=1$ term
of the far-zone series, i.e.\
\begin{equation}
F\,e^{-2i\epsilon\ln\epsilon}=1+i(2\gamma_E+2\ln2-\tfrac{37}{12})\epsilon+O(\epsilon^2).
\end{equation}
Combining, 
\begin{equation}
S^-=2i\pi F\,\left(\delta^2\Kc_+\over\epsilon\right)(\eta+\coth\pi\tau)[1+(\pi-i\nu_2)\epsilon+O(\epsilon^2)],
\end{equation}
which with
\begin{equation}
     \tau(1+\tau^2)(4+\tau^2)(1+\coth\pi\tau)=\Gamma(3+i\tau)\Gamma(3-i\tau){e^{\pi\tau}\over\pi}
\end{equation}
and $\pi(\tau+\epsilon)=(\omega-m\Omega_H)/2T_H$ gives the leading
exchange-odd amplitude, including its first relative correction:
\begin{equation}
S^-=-\frac{\epsilon^6\kappa^5}{900}\,\Gamma(3+i\tau)\Gamma(3-i\tau)
 e^{(\omega-m\Omega_H)/2T_H}
\times\left[1+i\epsilon\left(2\gamma_E+2\ln2-\frac{37}{12}-\nu_2\right)
 +\frac{5m}{9}q\epsilon+O(\epsilon^2)\right].
\label{eq:boltzmann}
\end{equation}
This form makes the thermal factor and the first subleading MST and far-zone
corrections explicit. Equivalently, the relative $O(\epsilon)$ correction can
be generated by the following operator acting on the leading $\tau$-dependent
factor (with $\partial_\tau$ differentiating that factor):
\begin{equation}
\frac1\kappa\partial_\tau+\pi+i\left(2\gamma_E+2\ln2-\frac{37}{12}-\nu_2\right)
+\frac{5m}{9}q,
\label{eq:eps7}
\end{equation}
whose constant and $q$-dependent terms are independently tested in
Sec.~\ref{supp:S7}. Acting on the non-polynomial part of Eq.~\eqref{eq:N6}, the
same operator also generates that of the order-$\epsilon^7$ coefficient in the
exchange-even sector; there this is a numerical result (Sec.~\ref{supp:S7}).
\item \emph{Exchange-even part.} There
$(1+\Xi)/2=1+i\pi\delta(\eta+\coth\pi\tau)+O(\delta^2)$ multiplies
$\Kc_+=(\delta^2\Kc_+)\delta^{-2}[1+\delta L+O(\delta^2)]$, with
$L=2\ln(2\epsilon\kappa)+\psi(3-i\tau)+\psi(-2-i\tau)+{\rm const}$. The finite
part is proportional to $L+i\pi(\eta+\coth\pi\tau)=2\ln(2\epsilon\kappa)+2\,{\rm
Re}\,\psi(3+i\tau)+i\pi\eta+{\rm const}$, with the normalization of
Eq.~\eqref{eq:even}. The pole term $(i/900)\epsilon^4\Pi_m/\nu_2$ of $H^+$
(checked numerically) is absent from $S$, whose $\epsilon^4$ coefficient is a
cubic polynomial in $q$: it cancels against $F$.
\end{enumerate}

\suppsection{supp:S7}{Numerical validation.}
The exchange law Eq.~\eqref{eq:law} agrees with MST to 60 digits for
$s=0,-1,-2$, $\ell\le3$, $q\in[0,0.9]\cup[1.5,3]$ and $M\omega\le0.3$.
Direct integration of the monodromy loops ($s=-2$, $\ell=m=2$, $q=0.6$,
$\epsilon=0.3$) verifies the two factorizations of $M_\infty$, the Floquet
eigenvalues and the lasso images to $10^{-8}$. The invariance of $\nu$ and of
the far-zone factor holds to $10^{-57}$ and $10^{-51}$.

For the low-frequency expansion, $B_6=i\Pi_m/450$ is verified to
$10^{-10}$; Eq.~\eqref{eq:odd} is exact for all $m$ and reproduces the
$q^6$ coefficients of Ref.~\cite{BjerrumBohr:2026kc}. Also
$[q^5]\mathcal C=16b_2$ to 20 digits and $\mathcal C/q^5\to64/15$.
The super-extremal $a^5$ coefficients agree with direct MST to $10^{-9}$.
The operator in Eq.~\eqref{eq:eps7} is verified in both sectors: its constant
terms agree to $3\times10^{-8}$, the coefficient $5m/9$ to $2\times10^{-8}$,
and no $q^2$ term is found above $10^{-9}$.

The leading-small-$M\omega$ coefficients from Eqs.~\eqref{eq:Hodd} and~\eqref{eq:R2m}
agree to $5\times10^{-6}$ for $s=-2$, $\ell=2$ and all $m$, and to
$2\times10^{-4}$ for the additional $\ell=3$ and $s=0$ cases. The $m=2$
super-extremal $a^6$ coefficients extrapolated over $300\le q\le5000$ are
$1.89648157$ and $1.89629630$, versus direct MST values $1.89648161$ and
$256/135$. The detailed expansion of Sec.~\ref{supp:S6} agrees with full MST
for $\epsilon\le4\times10^{-3}$: the $\mathcal X$-ratio coefficient to
$10^{-6}$, the $O(\epsilon)$ coefficient of $E$ to $10^{-5}$, and the
spheroidal coefficients $E_1,E_2$ to $10^{-9}$.

The unit-flux normalization (Sec.~\ref{supp:S8}) was tested for all $m$,
$0\le q\le0.95$, $0.01\le\epsilon\le1.2$, including superradiance and the
complex-$\nu$ branch. The Floquet-amplitude relations hold to $10^{-59}$,
the factorized and direct $\mathcal S_P$ agree to $10^{-60}$, and
$|\mathcal S_P|^2+\mathcal P_H=1$ to $2\times10^{-59}$. The exchange-odd
absorption formula holds for both $\eta$ to $2\times10^{-6}$; its fraction is
$1/2$ at $q=0$ and $\kappa/(1+\kappa)$ for $m=0$, to $5\times10^{-7}$.

\suppsection{supp:S8}{Flux balance and unit-flux normalization.}
For real-frequency gravitational perturbations we now fix the normalization
needed to turn the conserved radial identity into an energy flux. We use the
standard Kinnersley-tetrad, spin-$s=-2$ convention
\begin{equation}
 \rho^{-4}\psi_4=\frac{1}{\sqrt{2\pi}}
 \sum_{\ell m}\int d\omega\,
 e^{-i\omega t+im\phi}\,{}_{-2}S_{\ell m\omega}(\theta)
 R_{\ell m\omega}(r),\qquad
 \int_0^\pi |{}_{-2}S_{\ell m\omega}|^2\sin\theta\,d\theta=1.
 \label{eq:separation-flux-normalization}
\end{equation}
The physical ingoing solution has the future-horizon behavior
\begin{equation}
 R_{\ell m\omega}(r)\underset{r\to r_+}{\sim}
 Z^H_{\ell m\omega}\,\Delta^2e^{-ik_Hr_*},\qquad
 k_H=\omega-m\Omega_H.
 \label{eq:ZH-definition}
\end{equation}
Here $Z^H_{\ell m\omega}$ is the coefficient of the Kinnersley radial
function itself; it is not the regular-horizon Weyl scalar or the horizon
shear. The latter quantities are obtained from it by the
Teukolsky--Starobinsky identities and the Kinnersley-to-Hawking--Hartle
 tetrad transformation.

Let $\varkappa_H=2\pi T_H$ denote the (positive) surface gravity, to
 distinguish it from the dimensionless $\kappa=\sqrt{1-q^2}$ used in the
main text, and set $D_+=r_+^2+a^2=2Mr_+$. In the separation-constant
convention of the radial equation in the main text,
\begin{equation}
 \lambda=A_{-2,\ell m}(a\omega)+a^2\omega^2-2am\omega,
 \qquad \lambda\big|_{a\omega=0}=(\ell-1)(\ell+2),
\end{equation}
the squared gravitational Starobinsky constant is
\begin{align}
 |\mathcal C_{\ell m\omega}|^2={}&
 \big[(\lambda+2)^2+4am\omega-4a^2\omega^2\big]
 \big[\lambda^2+36am\omega-36a^2\omega^2\big]\notag\\
 &+(2\lambda+3)(96a^2\omega^2-48am\omega)
   +144\omega^2(M^2-a^2).
 \label{eq:starobinsky-norm}
\end{align}
The horizon-flux formula following from the shear/area balance and the
Teukolsky--Starobinsky conversion is then
\begin{equation}
 \mathcal F_H^{\ell m\omega}=
 \frac{128\,\omega k_H D_+^5}
      {|\mathcal C_{\ell m\omega}|^2}
 (k_H^2+\varkappa_H^2)(k_H^2+4\varkappa_H^2)
 |Z^H_{\ell m\omega}|^2 .
 \label{eq:explicit-horizon-flux}
\end{equation}
This is the standard Teukolsky horizon-absorption formula
\cite{TeukolskyPress:1974,Sasaki:2003xr}, integrated over the full sphere in
$G=c=1$ units with the Fourier normalization of
Eq.~\eqref{eq:separation-flux-normalization}. The factor
$1/\sqrt{2\pi}$ gives unit full-sphere norm, so no additional $2\pi$ occurs.
Restoring $G$ at fixed metric perturbation multiplies the flux by $1/G$. The
last two factors are the squared moduli of the horizon-frequency factors
$k_H+i\varkappa_H$ and $k_H+2i\varkappa_H$ generated by the four radial
Starobinsky derivatives; equivalently, they are
$(k_H^2+4\tilde\epsilon^2)(k_H^2+16\tilde\epsilon^2)$ for
$\tilde\epsilon=\varkappa_H/2$.

Thus, with $\mathcal A_H:=Z^H_{\ell m\omega}$, the normalization in
Eq.~\eqref{eq:physical-flux-balance} is explicitly
\begin{equation}
 \widetilde{\mathcal N}_{\ell m\omega}=
 \frac{128\,\omega D_+^5}
      {|\mathcal C_{\ell m\omega}|^2}
 (k_H^2+\varkappa_H^2)(k_H^2+4\varkappa_H^2)>0
 \quad(\omega>0).
 \label{eq:explicit-flux-normalization}
\end{equation}
The sign of $\mathcal F_H$ is therefore exactly the sign of $k_H$; it
vanishes at the superradiant threshold $\omega=m\Omega_H$.

The conservation law is the Teukolsky--Starobinsky cross-spin bilinear:
the spin-$+2$ and spin-$-2$ solutions are related by fourth-order Starobinsky
operators, and the bilinear is radius-independent. A same-spin Schr\"odinger
current is not appropriate because the separated Teukolsky potential is
generally complex. Evaluation at the horizon and infinity gives
\begin{equation}
 \mathcal F_{\rm in}^{\infty}-\mathcal F_{\rm out}^{\infty}
 =\mathcal F_H,
 \label{eq:physical-flux-balance}
\end{equation}
with, in the same normalization,
$\mathcal F_{\rm in}^{\infty}=128\,\omega^6|B^{\rm inc}_{-2,\ell m\omega}|^2/|\mathcal C_{\ell m\omega}|^2$
and $\mathcal F_{\rm out}^{\infty}=|B^{\rm ref}_{-2,\ell m\omega}|^2/(2\omega^2)$, where
$R_{\ell m\omega}\to B^{\rm inc}_{-2,\ell m\omega}r^{-1}e^{-i\omega r_*}
+B^{\rm ref}_{-2,\ell m\omega}r^{3}e^{i\omega r_*}$ at large $r$. Hence
\begin{equation}
 \mathcal P_H=\frac{\mathcal F_H}{\mathcal F_{\rm in}^{\infty}}
 =\frac{D_+^5\,k_H(k_H^2+\varkappa_H^2)(k_H^2+4\varkappa_H^2)}{\omega^5}
 \left|\frac{Z^H_{\ell m\omega}}{B^{\rm inc}_{-2,\ell m\omega}}\right|^2 .
 \label{eq:PH-ratio}
\end{equation}
For a unit-flux-normalized physical reflection coefficient
$\mathcal S_{\ell m\omega}$, this is equivalently
\begin{equation}
 \frac{\mathcal F_H}{\mathcal F_{\rm in}^{\infty}}
 =1-|\mathcal S_{\ell m\omega}|^2.
 \label{eq:absorption-probability}
\end{equation}
For the Kinnersley radial normalization above, the energy reflection
coefficient is
\begin{equation}
 \mathcal R_E=
 \frac{|\mathcal C_{\ell m\omega}|^2}{256\omega^8}
 \left|\frac{B^{\rm ref}_{-2,\ell m\omega}}
                  {B^{\rm inc}_{-2,\ell m\omega}}\right|^2,
 \qquad
 \mathcal P_H=1-\mathcal R_E .
 \label{eq:reflection-flux-conversion}
\end{equation}
This is the Teukolsky--Press flux conversion; the raw radial ratio is not
itself a unit-flux reflection amplitude. Equivalently, in the scattering
normalization used in the Kerr tidal-response literature,
\begin{equation}
 \mathcal S_{\ell m\omega}
 =e^{i\chi_{\ell m\omega}}
 \frac{\mathcal C_{\ell m\omega}}
      {16\omega^4}
 \frac{B^{\rm ref}_{-2,\ell m\omega}}
      {B^{\rm inc}_{-2,\ell m\omega}},
 \qquad
 \mathcal P_H=1-|\mathcal S_{\ell m\omega}|^2 .
 \label{eq:unit-flux-S}
\end{equation}
The MST factor $F_{\ell m}$ in Eq.~\eqref{eq:Smatrix} must consequently
be understood as the Starobinsky-normalized ratio of the incoming and
outgoing asymptotic coefficients of the $R_C^\nu$ Floquet solution:
\begin{equation}
 F_{\ell m}
 =e^{i\chi_{\ell m\omega}}
 \frac{\mathcal C_{\ell m\omega}}
      {16\omega^4}
 \frac{B^{\rm ref}[R_C^\nu]}
      {B^{\rm inc}[R_C^\nu]} .
 \label{eq:F-MST-matching}
\end{equation}
The coefficients in the last ratio are defined by expanding that same
Coulomb-wave MST solution at infinity in the $r^{-1}e^{-i\omega r_*}$
and $r^3e^{+i\omega r_*}$ branches. With this definition, the response
fraction in Eq.~\eqref{eq:Smatrix} is the remaining
$\Kc$-dependent factor in the unit-flux reflection coefficient.
Here $\mathcal C_{\ell m\omega}=D+12iPM\omega$ is the complex Starobinsky
constant for parity $P=\pm1$, with $D$ real and $|\mathcal C_{\ell m\omega}|^2$
given by Eq.~\eqref{eq:starobinsky-norm}. The helicity-preserving partial-wave
coefficient, which enters the Compton amplitude and was used for the expansions
of Sec.~\ref{supp:S4}, is the parity average and carries $D$ in place of
$\mathcal C_{\ell m\omega}$. It is not unit-flux normalized:
$|\mathcal S_P|^2=|S|^2+|S_{\rm flip}|^2$, with the helicity-reversing
coefficient $S_{\rm flip}=(12iM\omega/D)\,S$, so $1-|S|^2-\mathcal P_H\simeq\epsilon^2/16$.
Because $\mathcal C_{\ell m\omega}/D=1+iP\epsilon/4+O(\epsilon^2)$ is
exchange-even, Eqs.~\eqref{eq:odd} and~\eqref{eq:odd-absorption}, the
transcendental parts of Eqs.~\eqref{eq:N6} and~\eqref{eq:even}, the
fixed-$a\omega$, small-$M\omega$ limits and all quoted coefficients are the
same in both normalizations; only the
coefficients of degree below five of $p_5^{(m)}$ and the $O(\epsilon)$ phase
constant in Eqs.~\eqref{eq:boltzmann} and~\eqref{eq:eps7}, which shifts by $P/4$,
refer to the helicity-preserving normalization.

Equation~\eqref{eq:F-MST-matching} is checked by evaluating the two Floquet
amplitudes independently from the same continued-fraction coefficients. The
factorized $\mathcal S$ agrees with
$(\mathcal C/16\omega^4)B^{\rm ref}/B^{\rm inc}$ and satisfies
$|\mathcal S|^2+\mathcal P_H=1$ to 59 digits.

The decomposition $S_+=S^{\rm even}+S^{\rm odd}$ is nevertheless useful
for the physical outer-horizon solution. Once the unit-flux normalization
has been made, its contribution to the absorption probability is
\begin{equation}
 1-|S_+|^2=
 1-|S^{\rm even}|^2
 -2\operatorname{Re}\!\left[(S^{\rm even})^*S^{\rm odd}\right]
 -|S^{\rm odd}|^2.
 \label{eq:odd-flux-contribution}
\end{equation}
This is the precise amplitude-level sense in which the exchange-odd term
contributes to the physical flux balance. It is not a difference of two
horizon fluxes: the inner-horizon image is an auxiliary analytic solution,
not a second physical exterior scattering state. In particular, the
exchange-even part can also contribute to absorption, including the leading
superradiant flux. With the unit-flux $F_{\ell m}$, the exchange-odd term
of Eq.~\eqref{eq:odd-flux-contribution} agrees with
Eq.~\eqref{eq:odd-absorption} for both $\eta$ (with $\bar n\to\bar n+1$ for
$\eta=-1$); it is one half of $\mathcal P_H$ at $q=0$, and the fraction
$\kappa/(1+\kappa)$ of it for $m=0$ (Sec.~\ref{supp:S7}).

\end{document}